\documentclass[onecolumn,draft,prd,nofootinbib]{revtex4}
\usepackage{epsf}
\usepackage{dcolumn}
\usepackage{bm}

\newcommand{\be}{\begin{equation}}
\newcommand{\ee}{\end{equation}}

\def\ie{{\it i.e.},~}
\def\eg{{\it e.g.},~}
\def\etal{{\it et al.}~}
\def\4he{$^4$He}
\def\3he{$^3$He}
\def\7li{$^7$Li}
\def\Yp{Y$_{\rm P}$~}

\def\hii{H\thinspace{$\scriptstyle{\rm II}$}~}

\newcommand\la{\lower0.6ex\vbox{\hbox{\ensuremath{\buildrel{\textstyle<}\over{\sim}\ }}}}
\newcommand\ga{\lower0.6ex\vbox{\hbox{\ensuremath{\buildrel{\textstyle>}\over{\sim}\ }}}}

\begin{document}


\title{Cahill's Cosmological Model Exacerbates The Primordial 
Lithium Problem And Creates New Problems For Primordial 
Deuterium And Helium}

\author{Gary Steigman\\}

\affiliation{Departments of Physics and Astronomy\\
Center for Cosmology and Astro-Particle Physics\\
The Ohio State University, Columbus, OH 43210\\}

\date{{\today}\\}

\begin{abstract}

In a recent article R. T. Cahill claims that the cosmological model 
based on his ``new physics of a  dynamical 3-space" resolves the 
CMB-BBN \7li and \4he abundance anomalies.  In this note it is shown 
that this conclusion is wrong, resulting from a misunderstanding. 
In fact, primordial nucleosynthesis in this non-standard cosmological 
model {\bf exacerbates} the \7li problem and creates {\bf new problems} 
for primordial \4he and deuterium.

\end{abstract}


\maketitle

\section{~Introduction}
\label{intro}

In a series of papers \citep{cahill05,cahill09a,cahill09b,cahill09c} R. T. 
Cahill proposed a non-standard cosmological model based on his ``new 
physics of a dynamical 3-space".  In his most recent paper \citep{cahill09c} 
Cahill shows that this model predicts a {\bf slower} evolution of the early, 
radiation-dominated (RD) Universe and he {\bf misinterprets} this result to 
infer that the radiation (\eg the CMB photons) is {\bf hotter} than the baryons 
(nucleons), leading to an early-Universe ratio of baryons to photons which
is {\it different} from the present (and, compared to standard cosmologies).   
In fact, in his, as in other cosmologies, after the e$^{\pm}$ pairs have 
annihilated in the early Universe, prior to Big Bang Nucleosynthesis (BBN), 
the numbers of CMB photons and of baryons in every comoving volume are 
preserved and the ratio of baryons to photons, $\eta_{\rm B} \equiv 
n_{\rm B}/n_{\gamma}$, remains constant during the subsequent evolution 
of the Universe.  The baryon-to-photon ratio has nothing to do with how 
fast or slow the Universe is expanding.  In this note it is shown that 
since in Cahill's cosmology the early, Universe expands more slowly 
(compared to the standard cosmology), the predicted relic abundance 
of \7li {\bf increases}, while the predicted primordial abundances of 
D and \4he {\bf decrease}, in comparison to the predictions of standard 
BBN (SBBN).

In \S \ref{misunderstanding} the early, RD evolution of the Universe 
in Cahill's cosmology \cite{cahill09c} is described and the source of 
Cahill's misinterpretation of the implication of the model's slower 
evolution is identified.  In \S \ref{nsbbn} the primordial abundances 
of D, \4he, and \7li are calculated in this non-standard BBN (nSBBN) 
model and they are compared to the observationally-inferred relic 
abundances.  Our results are summarized and our conclusions presented 
in \S \ref{summary}.

\section{A Misunderstanding Of The Radiation-Dominated Evolution
of the Universe}
\label{misunderstanding}

According to Cahill \cite{cahill09c}, the expansion rate (the Hubble 
parameter, $H'$) of the early, RD Universe in his cosmological model 
is  related to the energy density in ``radiation" (\eg in all massless 
and/or extremely relativistic particles) by
\be
{H'(z)^{2} \over H_{0}^{2}} = {\Omega_{\rm R} \over 2}(1 + z)^{4},
\ee
where $\Omega_{\rm R}$ is the usual radiation density parameter and $z$ 
is the redshift.  It is the factor of $1/2$ in eq.~1 which distinguishes 
the early evolution of the Universe in this model from the standard model 
result: $H^{2}/H_{0}^{2} = \Omega_{\rm R}(1 + z)^{4}$.  In the standard 
model, as well as in Cahill's, the relation between the Hubble parameter 
and the age of the Universe during the RD epoch is: $Ht = 1/2$.  As a 
result, the age of the Universe at a fixed redshift (or temperature) 
in Cahill's model is {\bf larger} than in the standard cosmology: $t'/t 
= H/H' = \sqrt 2$.  Cahill's Universe expands more slowly than the 
standard cosmology (\ie it takes longer for the Universe to cool to a 
fixed temperature).  At a {\bf fixed time} after the beginning of the 
expansion, Cahill's Universe is {\bf hotter} than the standard cosmology.  
Cahill misinterprets this to draw the incorrect conclusion that in his 
model the photons are hotter than the baryons (resulting in a smaller 
baryon to photon ratio).  Independent of how fast the Universe expands, 
the numbers of CMB photons and of baryons (nucleons) in every comoving 
volume of the Universe are preserved (resulting from baryon number 
conservation for the baryons and from entropy conservation for the 
CMB photons).  The {\bf ratio} (by number) of baryons to photons is 
unchanged in Cahill's model (compared to the standard cosmology).

\section{BBN Abundances In A More Slowly Expanding Early Universe}
\label{nsbbn}

In Cahill's model, the early Universe expands more slowly than it 
does in the standard cosmology.  The early, RD evolution of the 
two models may be compared by comparing their expansion rates 
(Hubble parameters).  Their ratio, the expansion rate factor, 
$S \equiv H'/H$, introduced by Kneller \& Steigman (2004) \cite{ks}, 
is 
\be
S \equiv H'/H = 1/\sqrt{2}.
\ee
$S < 1$ corresponds to a more slowly evolving Universe.  In this 
case there is more time to bridge the gap at mass-5 and build the 
primordial abundance of \7li, {\bf increasing} (not decreasing) 
its relic abundance compared to the prediction of SBBN ($S = 1$).  
The slower expansion also allows more time to burn deuterium 
to \3he and \4he, {\bf reducing} the relic D abundance and, with 
more time available for neutrons to decay, the primordial \4he 
mass fraction is {\bf reduced} too.
 
Kneller \& Steigman (2004) \cite{ks} provide simple fitting formulae 
for the BBN light element yields when $S \neq 1$, and these have 
been updated in Steigman (2007) \cite {steigman07}.  Although the
very small value of $S$ for Cahill's model is somewhat outside of 
the range of values where the fits are as accurate as quoted in 
references \cite{ks,steigman07}, they still provide sufficiently 
accurate quantitative estimates for the relic abundances in this 
non-standard BBN (nSBBN) model.\\ 

\underline{Deuterium}

For deuterium the nSBBN-predicted relic abundance \cite{steigman07} 
is $y_{\rm DP}\equiv 10^{5}$(D/H)$_{\rm P} = 1.7$ (or, log $y_{\rm 
DP} = 0.22$), to be compared with the observationally-inferred 
value \cite{pettini} of $y_{\rm DP} = 2.8$ (or, log $y_{\rm DP} = 
0.45$).  The primordial D abundance predicted by Cahill's model 
is {\bf too low} by a factor of $\sim 1.7$.\\

\underline{Helium-4}

Unquantified (and often unidentified) systematic errors plague 
the \hii region observational data used to infer the primordial 
abundance of \4he (see, \eg Steigman (2007) \cite{steigman07} for a 
review).  Given this state of affairs, a few good (well observed 
and analyzed) \hii regions may be of more value than hundreds of 
\hii regions with lower quality data and analysis (see, \eg the 
discussion in Steigman (2009) \cite{steigman09}).  For example, 
from the analysis of Olive \& Skillman \cite{os}, \Yp $\leq 0.250 
\pm 0.003$, while from similar data set, analyzed carefully by 
Peimbert, Luridiana, \& Peimbert \cite{plp}, \Yp $\leq 0.252 \pm 
0.004$.  These {\it upper bounds} to the \4he mass fraction, with 
their more realistic errors, are consistent with the estimate, 
\Yp $= 0.247 \pm 0.001$ (or, using new emissivities, \Yp $= 0.252 
\pm 0.001$) from Izotov, Thuan, \& Stasinska (2007) \cite{its}.  
In stark contrast, in Cahill's model the \4he nSBBN-predicted 
relic mass fraction is \Yp = 0.202.  The cosmological model 
proposed by Cahill creates a new, very serious \4he anomaly.\\

\underline{Lithium-7}

The lithium abundances derived from observations of the very 
most metal-poor halo and globular cluster stars in the Galaxy 
\cite{boesgaard05,asplund06,aoki09,lind09} lie well below the 
SBBN-predicted value ([Li]$_{\rm P,SBBN} \equiv 12 + 
$log(Li/H)$_{\rm P,SBBN} \approx 2.7$).  The discrepancy between 
the predictions and the observations is a factor of $\sim 3 - 
5$ (see, \eg \cite{steigman07,steigman09} and references therein).  
In his most recent paper \cite{cahill09c}, Cahill claims that his 
new cosmological model resolves the lithium abundance anomaly.  
In fact, for this nSBBN model, the predicted relic lithium 
abundance, [Li]$_{\rm P} \approx 2.8$, is $\sim 30\%$ larger 
than for SBBN, exacerbating, not resolving, the primordial 
lithium problem.\\

Because of the slower early evolution of the Universe 
in this model, the BBN-predicted abundances of D and \4he are 
too small and that of \7li is too large, to be consistent with 
the observational data.  The relic abundances of D, \4he, and 
\7li provide the nails in the coffin of this model.  

\section{Summary And Conclusions}
\label{summary}

During its early evolution the cosmological model proposed by 
Cahill \cite{cahill05,cahill09a,cahill09b,cahill09c} evolves more 
slowly than does the standard cosmological model.  In this model 
the slow, early evolution results in relic D and \4he abundances 
far too small to be consistent with the relic abundances inferred 
from the observational data.  And, with more time available to 
synthesize \7li, the \7li relic abundance is even larger than 
that predicted by SBBN, thus exacerbating the lithium problem.  
As a consequence of these conflicts it must be conluded that 
the cosmological model proposed by Cahill is incompatible with
the observationally-inferred abundances of the light nuclides
produced during BBN.  Cahill's model is not a viable cosmological
model.

\acknowledgments

The research of GS is supported at The Ohio State University 
by a grant from the US Department of Energy.  I am pleased to
acknowledge an informative exchange of emails with R.~T.~Cahill.

\end{document}